\documentclass{aa}

\usepackage{natbib}
\bibpunct{(}{)}{;}{a}{}{,} 
\usepackage{color}
\usepackage{graphicx}
\usepackage[varg]{txfonts}

\begin{document}

\renewcommand{\vec}[1]{\bmath{#1}}
\newcommand{\sign}{\textrm{sign}}
\newcommand{\pd}[2]{\frac{\partial #1}{\partial #2}}
\newcommand{\DS}{\displaystyle}
\newcommand{\HALF}{\frac{1}{2}}
\newcommand{\mathi}{\rm i}

\renewcommand{\vec}[1]{\mathbf{#1}}
\newcommand{\hvec}[1]{\hat{\mathbf{#1}}}
\newcommand{\av}[1]{\left<#1\right>}
\newcommand{\red}[1]{\color{red} #1 \color{black}}
\newcommand{\blue}[1]{\color{blue} #1 \color{black}}

\title{Recollimation shocks and radiative losses in extragalactic relativistic jets }
\author
       {G. Bodo$^1$ \&  F. Tavecchio$^2$}

\institute{INAF -- Osservatorio Astrofisico di Torino, Strada Osservatorio 20, I--10025 Pino Torinese, Italy
\and  INAF -- Osservatorio Astronomico di Brera, via E. Bianchi 46, I--23807 Merate, Italy}
\titlerunning{Recollimation shocks in jets.}
\authorrunning{Bodo \& Tavecchio}
\date{Accepted ??. Received ??; in original form ??}
\abstract{We present the results of state-of-the-art simulations of recollimation shocks induced by the interaction of a relativistic jet with an external medium, including the effect of radiative losses of the shocked gas. Our simulations confirm that -- as suggested by earlier semi-analytical models -- the post-shock pressure loss induced by radiative losses may lead to a stationary equilibrium state characterized by a very strong {\it focusing} of the flow, with the formation of quite narrow nozzles, with cross-sectional radii as small as $10^{-3}$ times the length scale of the jet. We also study the time-dependent evolution of the jet structure induced  of a density perturbation injected at the flow base. The set-up and the results of the simulations are particularly relevant for the interpretation of the observed rapid variability of the $\gamma$-ray emission associated to flat spectrum radio quasars. In particular, the combined effects of jet focusing and Doppler beaming of the observed radiation make it possible to explain the sub-hour flaring events such as that observed in the FSRQ PKS 1222+216 by MAGIC.}
\maketitle

\section{Introduction}
%
%
%

Extragalactic relativistic jets -- collimated outflows of plasma expelled by supermassive black holes residing in central regions of active galactic nuclei and traveling for hundreds of kpc -- are among the most fascinating astrophysical structures. Their phenomenology is regulated by complex time-dependent physical processes involving magnetic fields and plasma. Shocks and/or magnetic reconnection sites are able to accelerate charged particles to ultrarelativistic energies, whose presence is flagged by intense non-thermal emission over the entire electromagnetic spectrum, up to the TeV band \citep[e.g.][for a review]{Romero17}. Relativistic jets are best studied in blazars \citep[see e.g.][]{Urry95}. The jet of these sources is closely aligned toward the line of sight and, thanks to this favorable geometry, their non-thermal emission is strongly amplified by the relativistic beaming effects. 

 The observation of the emission from blazar jets at the highest energies, accessible thanks to space ({\it Fermi}) and ground based (Cherenkov arrays) instruments, is revealing numerous unexpected features. One of the most intriguing and challenging aspects is the ultra-fast variability (doubling time $\lesssim1$ hour, down to few minutes) detected in several blazars \citep{Aharonian07, Albert07, Arlen13, Aleksic11, Aleksic14}. These very small timescale are often smaller than the light crossing time of the horizon of the supermassive black hole \citep[e.g.][]{Vovk15} and their interpretation requires extreme physical conditions \citep[e.g.][]{Begelman08}. In particular, very rapid variability events detected in the most powerful blazars, the Flat Spectrum Radio Quasars (FSRQ), are the most challenging to interpret. 
FSRQ display the classical broad emission lines of quasars, flagging the existence of a broad line region in which clouds of   photoionized gas reprocessed part of the intense continuum emitted by the accretion flow. Due to the anticipated absorption of the $\gamma$-ray radiation above few tens of GeV through the interaction with soft (UV) photons \citep[e.g.][]{Liu06}, one can put robust lower limits to the distance of the emitting region from the central black hole since, to avoid absorption, the emission is constrained to occur beyond the BLR radius $R_{\rm BLR}$, of the order of $\sim0.1$ pc. For a conical geometry of the jet (with semiaperture $\theta_{\rm j}\approx 0.1$) this lower limit to the distance readily translates into a lower limit for the jet radius and (through light crossing time argument) to the minimum variability timescale, $\Delta t \gtrsim R_{\rm BLR} \theta_{\rm j}/c\mathcal{D}=10^5 \; \mathcal{D}^{-1}_1$ s (for which we use a Doppler factor\footnote{defined as $\mathcal{D}=[\gamma_{\rm j}(1-\beta_{\rm j}\cos \theta_{\rm v})]^{-1}$, where $\gamma_{\rm j}$ is the jet Lorentz factor, $\beta_{\rm j}$ the jet speed (in units of the speed of light $c$) and $\theta_{\rm v}$ the viewing angle.} $\mathcal{D}=10$), a value clearly not compatible with the most extreme minute-timescale events. One of the first FSRQ showing rapid variability events was PKS 1222+122 which, during a flare in June 2010, was detected by the MAGIC telescopes with a flux increasing with a doubling timescale of about 10 minutes \citep{Aleksic11}. Since the discovery of PKS 1222+122 other FSRQ have been observed to vary on short timescales. Recently,  3C 279 has been detected during a pointed {\it Fermi}/LAT observation to vary with a timescale of about 5 minutes \citep{Ackermann16}.

In the framework of the classical shock-in-jet scenario for particle acceleration \citep[e.g.][]{Aller85} is clearly hard to explain very short flares. At a first sight, the only possibility to explain the fast variability in FSRQ seems to admit the existence of compact ($R\lesssim 10^{15}$ cm) emission regions embedded in the jet at distances of the order of $\sim$0.1 pc from the central engine \citep[e.g.][]{Ghisellini08, Tavecchio11}. Generally, such compact regions have been identified in active {\it substructures} of the jet, such as plasmoids resulting from efficient reconnection of the magnetic field \citep{Giannios13, Petropoulou16} or in turbulent cells embedded in the relativistic flow \citep{Narayan12, Marscher14}. These scenarios have been widely discussed in the past \citep[see also][]{Aharonian17}.
However, one can adopt a different view and explore the possibility that the emission occurs within  the entire (or a large fraction of the) jet. This could be reconciled with the fast variability if the jet suffers a strong {\it recollimation} interacting with an external medium. Jet recollimation has been widely studied in the past \citep[e.g.][]{Komissarov98, Sokolov04, Stawarz06, Nalewajko09}. \citet{Bromberg09} extended the calculations including the effect of strong radiative losses of the shocked plasma. The cooling, causing a loss of pressure in the compressed region, allows the jet to be ``squeezed" by large factors. Bromberg \& Levinson applied their scenario to the case of BL Lac objects and radiogalaxies but -- as proposed in \citet{Tavecchio11}  -- it can also be applied to FSRQ, for which the cooling of the plasma (dominated by the inverse Compton scattering of the ambient soft photons \citep[e.g.][]{Ghisellini98} is expected to be even more severe than for other sources. 

In this paper we intend to study the recollimation of a jet subject to important radiative losses extending the previous approximated semi-analytical approach of \citet{Bromberg09},  by performing high resolution axisymmetric numerical simulations using of the state-of-the-art  numerical code PLUTO \citep{PLUTO, AMRPLUTO}. The structure of the paper is the following: in Sect. 2 we illustrate the model and the setup used. In Sect. 3 we show the results of simulations for stationary states and  for the time-dependent evolution of a density perturbation. In Sect. 4 we conclude.

\section{The model}
%
%
%
 We want to construct a numerical model of the confinement of a relativistic jet by the pressure of the surrounding medium.  We assume axisymmetry and make use of the cylindrical coordinates $r$ and $z$, where $z$ is the coordinate along the jet axis and $r$ is the radial coordinate. We inject a conical relativistic jet, with opening angle $\theta_0$,  in an overpressured confining region with a pressure profile
 \begin{equation}
p_{ext} (z) = p_a  \left( \frac{z}{z_0} \right)^{-2}
\end{equation}
and a density profile
\begin{equation}
\rho_{ext} = \rho_a \left( \frac{z}{z_0} \right)^{-2}
\end{equation}
where $z_0$ represents the height at which the jet starts to interact with the confining region.  

The jet is injected in the confining region ($z_0=1$) with a Lorentz factor $\gamma_j$ and a radius $r_j = \theta_0 z_0$, and is subject to radiative losses by non thermal processes.   In this paper we don't consider the dynamical effects of a magnetic field, therefore the  evolution  equations determining the jet dynamics are the continuity equation 
\begin{equation}\label{eq:continuity}
\partial_\nu (\rho \vec u^\nu) = 0,
\end{equation}
where $\rho$ is the proper density and $u^\mu$ is the four-velocity, and the energy momentum conservation
\begin{equation} \label{eq:conservation}
  \partial_{\nu} T_{}^{{\mu} \nu} = S^{{\mu}} 
  \end{equation}
where $T_{}^{{\mu} \nu} $ is the stress-energy tensor of a perfect gas
\begin{equation}
T_{}^{{\mu} \nu} = w u^\mu u^\nu + p g^{{\mu} \nu} 
\end{equation}
where  $w$ and $p$ denote, respectively, the proper enthalpy and pressure and $g^{{\mu} \nu} $ is the metric tensor for a flat space. $S^{\mu}$ denotes a source term associated with radiative losses, whose form is specified below.   The system of equations (\ref{eq:continuity}) and (\ref{eq:conservation}) 
is completed by providing an equation of state relating $w$, $\rho$ and $p$. 
Following \cite{Mignone05b}, we adopt the following prescription: 
\begin{equation} \label{eq:eos}
w = \frac{5}{2}p + \sqrt{\frac{9}{4}p^2 + \rho^2}
\end{equation}
which closely reproduces the thermodynamics of the Synge gas for a 
single-specie relativistic perfect fluid with a smooth transition  from the adiabatic exponent $\Gamma = 5/3$ in the non relativistic limit to
$\Gamma = 4/3$ in the ultra relativistic case (here and in the following we always put $c = 1$). 

For radiative losses we assume a very simple form, they are taken to be proportional to the pressure and we additionally assume that it is only the gas above a certain threshold temperature $T_c$ that radiates. The first assumption is widely adopted \citep[e.g.][]{Komissarov98, Bromberg09}. For the case under study here the cooling is dominated by the inverse Compton emission on a fixed target radiation field. The emissivity (the frequency integrated radiated power per unit volume) of the non-thermal particles measured in the  plasma comoving frame, $\epsilon\equiv - S^{0}$, can be expressed as:
\begin{equation}
\epsilon=\frac{4}{3}\sigma_T c \, U_{\rm ext} \int_{\gamma_{\rm min}}^{\gamma_{\rm max}} n(\gamma) \gamma^2 d\gamma
\end{equation}
where $\sigma_T$ is the Thomson cross section, $U_{\rm ext}$ is the energy density of the external radiation field, $\gamma$ is the electron Lorentz factor and $n(\gamma)$ is the non-thermal electron energy distribution, for simplicity assumed to be a power law with slope 2, $n(\gamma)=k\gamma^{-2}$, in the range $\gamma_{\rm min} < \gamma < \gamma_{\rm max}$, with $\gamma_{\rm min}\ll \gamma_{\rm max}$ (all quantities are expressed in the flow frame). Performing the integral one obtains $\epsilon=(4/3)\sigma_T c \, U_{\rm ext} k \gamma_{\rm max}$. The energy density of the non-thermal relativistic electrons is:
\begin{equation}
U_{\rm e}=m_e c^2 \int_{\gamma_{\rm min}}^{\gamma_{\rm max}} n(\gamma) \gamma d\gamma =k m_e c^2 \Lambda,
\end{equation}
where $\Lambda=\ln(\gamma_{\rm max}/\gamma_{\rm min})$. Combining the two equations above we can write:
\begin{equation}
\epsilon=\frac{4\sigma_T}{3m_ec} \frac{\gamma_{\rm max}}{\Lambda}\, U_{\rm ext} U_{\rm e} = \frac{p_{\rm e}}{\tau_c}
\end{equation}
where in the last step we used $p_{\rm e}=U_{\rm e}/3$ (valid for ultrarelativistic electrons) and we have defined an effective cooling time:
\begin{equation}
{\tau_c}=\frac{m_ec \,\Lambda}{4\sigma_T \gamma_{\rm max} U_{\rm ext}}.
\end{equation}
The pressure of the non-thermal particles is supposed to be a fraction $\xi_{\rm e}$ of the thermal one, i.e $p_{\rm e}= \xi_{\rm e} p$.
We note that the relation $\epsilon\propto p_{\rm e}$ is valid if, as assumed here, the electron losses are determined by the inverse Compton emission on a fixed target radiation field. In case the losses are dominated by synchrotron or synchrotron-self Compton emission one should also consider the role of the magnetic field. The assumption on the critical temperature is intended to mimic the idea that high temperatures flag the presence of relativistic particles in the flow. A last point concerns the fact that (as also pointed out by \citet{Nalewajko09}), in a realistic treatment one should assume that relativistic non-thermal particles are accelerated close to the shock fronts and then advected in the other regions of the jet. Both the electron distribution and $\xi_{\rm e}$ should thus be treated as functions of the position in the flow. For simplicity this is not considered here. In any case, we expect that the global dynamical effects of cooling, such as those explored here, do not dramatically depends on these details. Similarly to \citet{Bromberg09} the quantities characterizing the non-thermal population are assumed constant in the jet. Implicitly this choice assumes the existence of some mechanism (e.g. turbulence) continuously supplying the energy lost by particles through the emission.

With the assumptions above, and assuming that, in the rest frame of the gas, radiation is isotropic (for our applications we can safely neglect the anisotropy of the external Compton radiation field, \citet{Dermer95, Ghisellini10}),  we  can express  $S^\mu$ as
\begin{equation}\label{eq:radloss}
S^\mu = (-\frac{\xi_{\rm e}p}{\tau_c}, 0, 0, 0).
\end{equation}
In the laboratory frame we can then write
\begin{equation}
S^\mu = -\frac{\xi_{\rm e}p}{\tau_c} u^\mu
\end{equation}
 
Equations (\ref{eq:continuity}) and (\ref{eq:conservation}) are solved numerically by using the adaptive mesh refinenment (AMR) version of the PLUTO code \citep{PLUTO, AMRPLUTO}, with a second order scheme and HLLC  Riemann solver \citep{Mignone05a}. We perform two-dimensional axisymmetric simulations using the cylindrical coordinates $r$ and $z$ on a grid that covers the domain $0 < r < r_D$, $1 < z < z_D$, where $r_D = 1.5 z_0$ and $z_D = 9 z_0$. We make use of 6 levels of refinement, the base grid is made by $64 \times 384$ points  and we have an equivalent maximum resolution of $4096 \times 24576$ points. The boundary conditions are reflective on the axis $r=0$, outflow at the outer boundaries $z_D=9z_0$ and $r_D= 1.5 z_0$. At the lower boundary $z_0 = 1$, for $r < r_j$ we have inflow conditions injecting the jet flow, while, for $r > r_j$, we have reflective conditions. 

The jet is injected with a proper density $\rho_j =  10^{-6}\rho_a$, where $\rho_a$ is the ambient density at $z=z_0$, with a Lorentz factor  $\gamma_j = 10$ and and  pressure is  $p_j = 7 \times 10^{-3} p_a$.  The corresponding energy flux is: 
\begin{equation}
L_j = \pi \theta_0^2 z_0^2 w_j \gamma_j^2 v_j
\end{equation}
where $w_j$, according to Eq. (\ref{eq:eos}) is
\begin{equation}
w_j = \frac{5}{2}p_j + \sqrt{\frac{9}{4}p_j^2 + \rho_j^2}
\end{equation}

In the simulations we fix the parameters to values representative for the flaring state of PKS 1222+216 \citep[e.g.][]{Tavecchio11}. For the energy flux we use $L_j=10^{46}$ erg s$^{-1}$, Using $\gamma_{\rm max}=10^5$, $\Lambda\simeq 10$ (quite insensitive on the exact value of $\gamma_{\rm min}$) and $U_{\rm ext}=3\times 10^{-2}$ (suitable to reproduce the IR radiation field of the torus \citep[e.g.][]{Ghisellini08} we obtain $\tau_c\simeq 3\times 10^4$ s. Since we assume that the jet is already beyond the BLR we fix $z_0$ to the BLR radius, $z_0=7\times 10^{17}$ cm. The cooling time is thus of the order of $\tau_c=10^{-3} z_0/c$. We investigate the effects of radiative losses by performing a series of simulations with different values of $\xi_{\rm e}$ in the range $0.01-0.1$.

\section{Results}
%
%
%

\subsection{The equilibrium configuration}
\label{sect:equil}

\begin{figure}[b]
   \centering
\includegraphics[width=9cm]{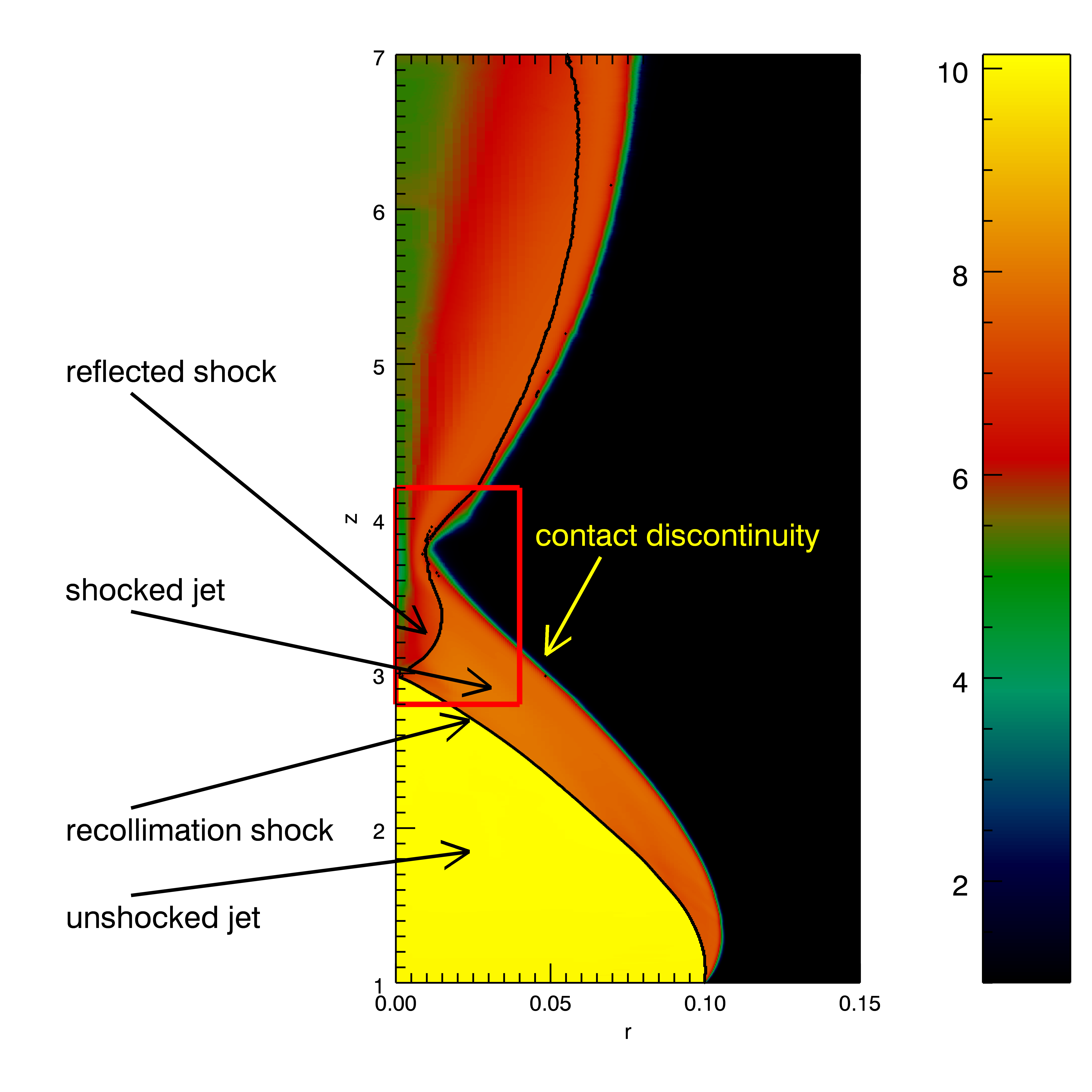} 
   \caption{\small  Distribution of the Lorentz factor for the case with $\xi_{\rm e}=0.01$, showing the main features of the steady solution. The red rectangle individuates the region shown in Fig. \ref{fig:zoom}. Notice that the radial scale is strongly stretched.}
\label{fig:lorentz}
\end{figure}
\begin{figure*}[t!]
   \centering
   \includegraphics[width=15cm]{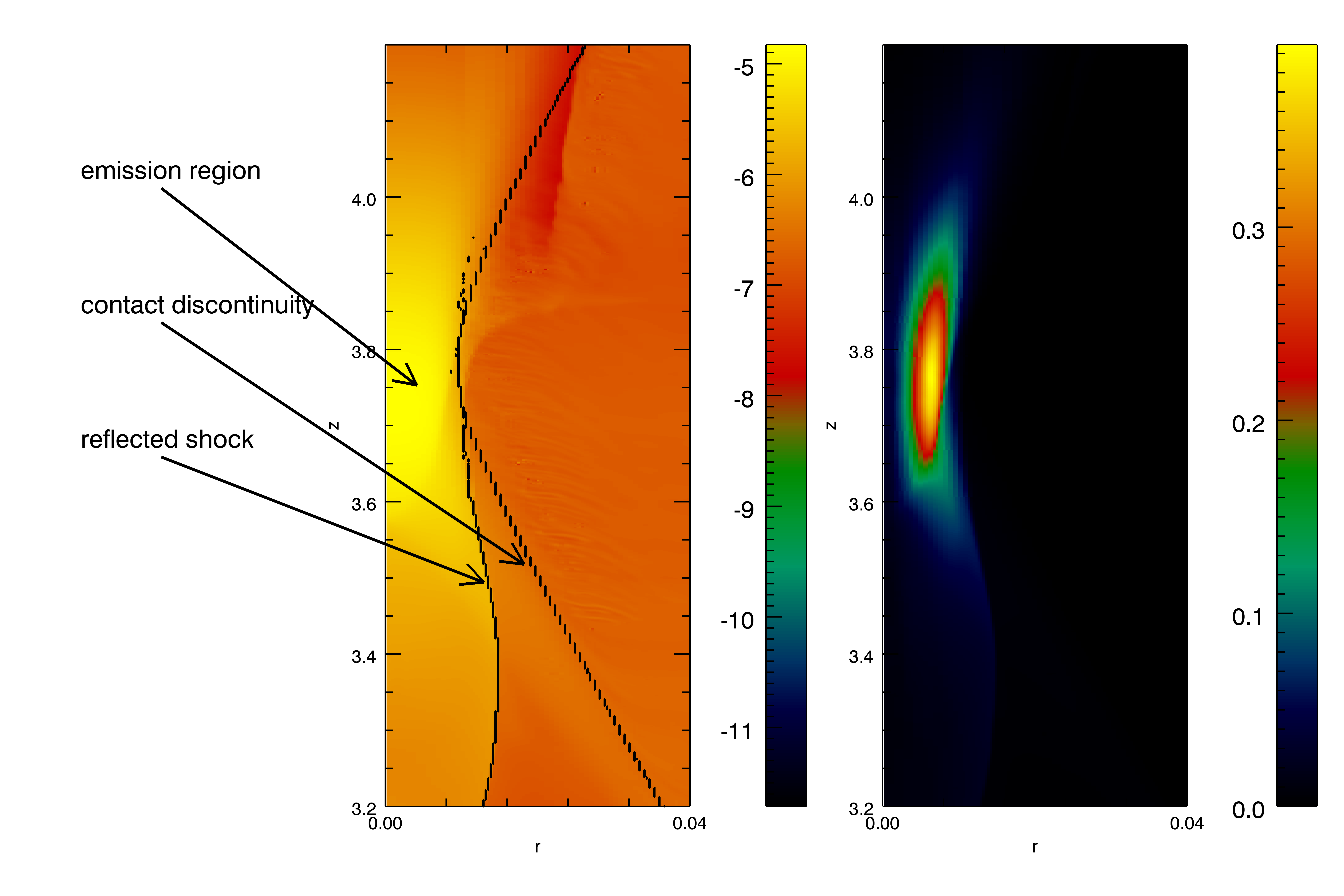} 
   \caption{\small  The left panel shows the pressure distribution in the region marked by the red rectangle in Fig. \ref{fig:lorentz} for the case with $\xi_{\rm e}=0.01$. The arrows indicate the main features already displayed in Fig. \ref{fig:lorentz}. The right panel displays the distribution of the observed emissivity (see Eq. (\ref{eq:emisobs}) }
   \label{fig:zoom}
\end{figure*}

 The characteristics of recollimation shocks resulting from the interaction of a under-pressured relativistic jet with external confining material have been discussed in the past in particular by \citet{Komissarov98, Bromberg07, Bromberg09, Nalewajko09}. The gross structure consists of a contact discontinuity separating the shocked jet layer and the ambient medium and a  recollimation shock. Komissarov \& Falle (1997) derived, under suitable approximations, simple analytical formulae for the geometrical properties of the shock. Assuming a power law profile for the pressure of the external gas, $p_{\rm ext}(z)=az^{-\eta}$, the profile $r(z)$ of the reconfinement shock follows the differential equation:
\begin{equation}
\frac{dr}{dz}=\frac{r}{z} -Az^{\delta}
\end{equation}
where $\delta=1-\eta/2$ and $A=(\pi a\theta^2_0 c/\mu\beta_jL_j)^{1/2}$, where $\mu=0.7$.
The solution of this equation, with the condition $r=r_0$ at $z=z_0$ is:
\begin{equation}
r(z)=z\theta_j - \frac{A}{\delta}z \theta_j \left[z^{\delta} -z_0^{\delta}\right],
\end{equation}
for $\eta\neq 2$, and $r(z)=zA\ln(z_c/z)$, for $\eta=2$ (i.e. $\delta=0$) where $z_c=z_0 \exp(r_0/z_0A)$. For the contact discontinuity surface a similar analytical expression can be derived (Bromberg \& Levinson 2007), $r_c(z)=r_0(z/z_0)^{\eta/4}$.

 %
\begin{figure*}[t] 
   \centering
   \includegraphics[width=10cm]{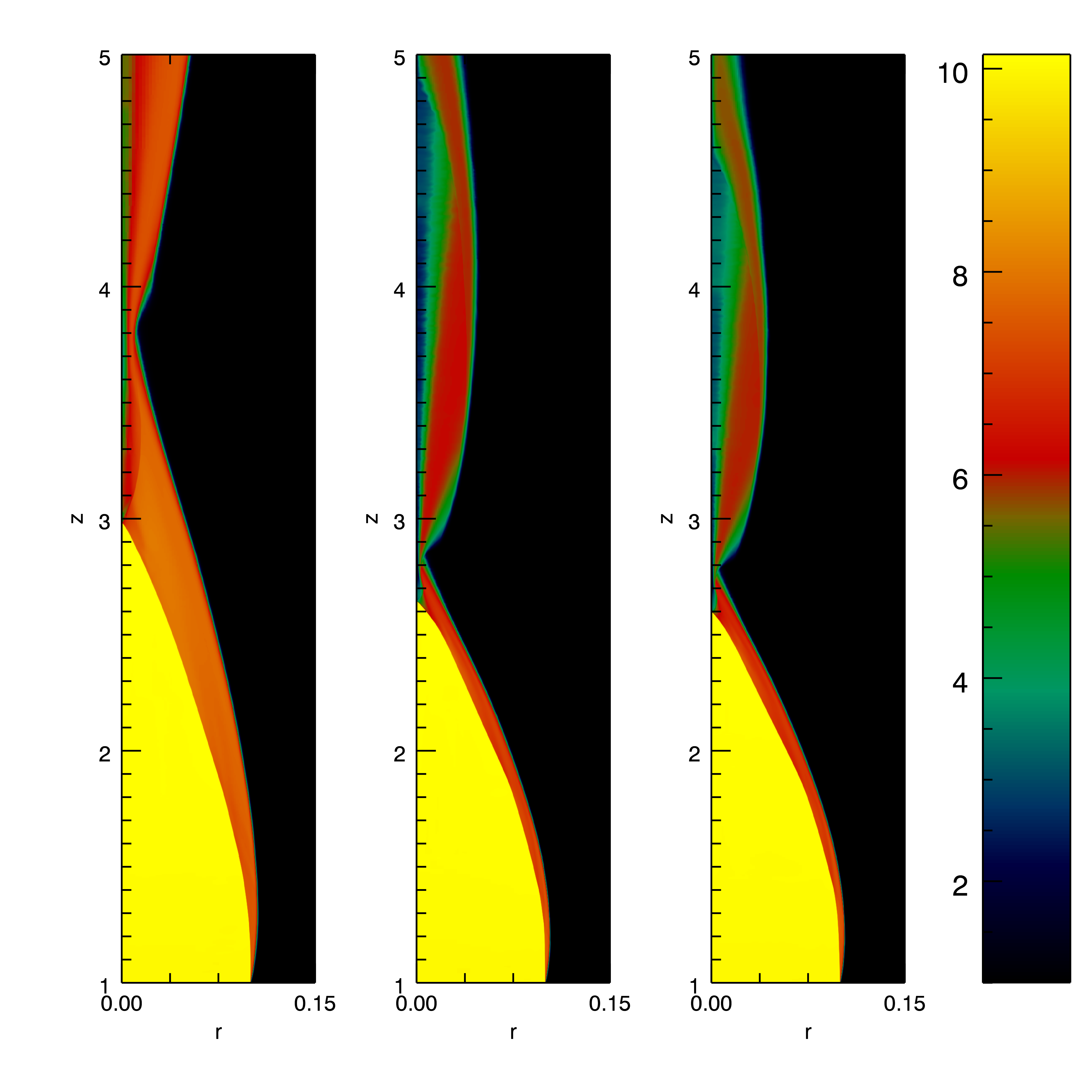} 
   \caption{\small Distribution of the Lorentz factors for the three cases with different values of $\xi_{\rm e}$, i.e. $\xi_{\rm e}=0.01$ (left panel), $\xi_{\rm e}=0.03$ (middle panel) and $\xi_{\rm e}=0.05$ (right panel).}
   \label{fig:lorentz3}
\end{figure*}

These analytical solutions do not include the possible cooling of the post-shock plasma. \citet{Bromberg07} and \citet{Bromberg09} constructed a class of semianalytical models for the confinment of the jet by the external pressure including also the effect of the pressure loss due to radiation losses. As above, these solutions are subject to some approximations and limitations:  in the shocked layer the flow parameters are assumed to depend only on the  coordinate along the jet axis $z$ (gradients along $r$ have been included by \citet{Nalewajko09} and, more critically, they do not get the structure of the solutions for values of $z > z^*$, where $z^*$ is the point where the recollimation shock reaches the jet axis.

Here we use a different approach, by which we overcome the two limitations mentioned above, solving numerically the time dependent equations and following the jet evolution until it reaches a steady configuration.  We start the simulations at $z=z_0$ with a conical jet of opening angle $\theta_0$ established over the entire computational domain. As the simulation evolves in time, the jet is compressed by the higher pressure of the ambient medium, the recollimation shock is formed and the system evolves towards its steady structure, that is reached at about   $t=300 z_0/c$. The main features of the steady solutions mentioned above can be observed in Fig. \ref{fig:lorentz}, where we show the Lorentz factor distribution for the case with the smaller value of $\xi_{\rm e} = 10^{-2}$. In the figure we see the region of unshocked jet material separated by the recollimation shock from the shocked jet layer.  A contact discontinuity separates this shocked material from the external medium, which appears in black. The conical recollimation shock reaches the jet axis at  $z=z^* \sim 3 z_0$, where it is reflected and gives rise to a strong jet deceleration. From the detailed view of the pressure distribution, shown in the left panel of Fig. \ref{fig:zoom}, we can observe the formation of a complex shock structure, which leads to a pressure maximum for $3.6 z_0 < z < 3.8 z_0$ .  Having linked the emissivity of non-thermal particles to the value of the pressure (Eq. \ref{eq:radloss}), the region of maximum pressure corresponds also to a maximum of radiative losses. Therefore, in the region around the axis, the jet, having lost a fraction of its energy flux by radiation, will remain at a Lorentz factor quite lower than its injection value (Fig. \ref{fig:lorentz}). This slower region on the axis is surrounded by a faster region, formed by material that has suffered less radiative losses.    
 
Considering a jet propagating towards the observer with viewing angle $\theta_{\rm v}=1/\gamma_{\rm j}$, the Doppler factor will be $\mathcal{D}=\gamma_{\rm j}$ and, with the assumptions described above, the {\it apparent} (Doppler boosted) emissivity of each fluid element as measured by the observer is given by:
\begin{equation} \label{eq:emisobs}
\epsilon _{\rm obs}=  \frac{p_{\rm e}}{\tau_c} \, \mathcal{D}^3.
\end{equation}
The location of the emissivity maximum depends both on the distribution of pressure and on the distribution of Lorentz factor within the flow. Since the Lorentz factor is lower on the axis and increases away from it, we expect to find the maximum of $\epsilon_{\rm obs}$ away from the axis. In fact from the right panel of Fig.  \ref{fig:zoom}, where we display the distribution of the observed emissivity, we can see that most of the observed luminosity originates from an elongated region at some distance from the axis, where the product of the pressure and the Lorentz factor is maximized. We remark  that, although the thickness of this region is very small ($<0.01 \,z_0$), its length is of the order of $\approx 0.2\,z_0$.

The effects of increasing the radiative losses are illustrated  by  Fig. \ref{fig:lorentz3}, where we show the distributions of the Lorentz factor for three cases with different values of $\xi_{\rm e}$.  From the figure we can observe a progressive strong reduction in transverse size of the region of shocked jet material, between the recollimation shock and the contact discontinuity, for increasing $\xi_{\rm e}$. The ``squeezing" of the jet also becomes more effective and the position of the minimum jet transverse radius comes closer to the point where the recollimation shock reaches the axis.  The Lorentz factor on the axis is also more strongly decreased because of the increased energy losses. Note also that after a moderate re-expansion, following the development of the reflection shock, the jet is recollimated again.

\begin{figure*}[t]  \label{fig:prem}
   \centering
   \includegraphics[width=10cm]{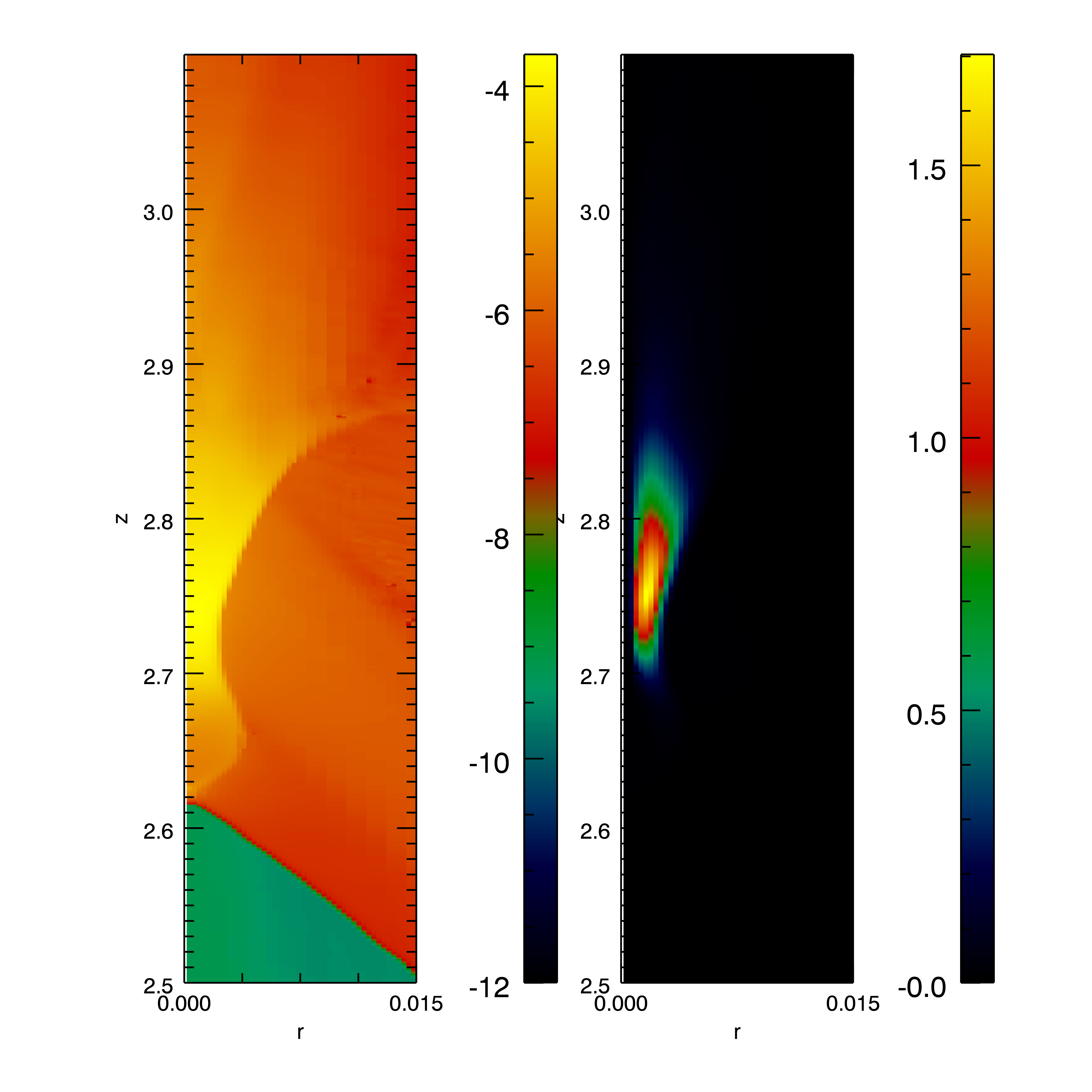} 
   \caption{\small The left panel shows the pressure distribution in the region marked by the red rectangle in Fig. \ref{fig:lorentz} for the case with $\xi_{\rm e}=0.05$. The right panel displays the distribution of the emissivity in the observer frame (see Eq. (\ref{eq:emisobs} ) }
\end{figure*}

Fig. \ref{fig:prem} shows a zoomed view of the distributions of  pressure and observed emissivity for the case with $\xi_{\rm e}=0.05$. The pressure distribution shows an overall shape similar to the previous case, shown in Fig. \ref{fig:zoom}, however it is much reduced in size and the maximum value is about one order of magnitude larger. Consequently, also the emission has a similar shape, with much reduced size and higher peak value.

\begin{figure}[h!]
   \centering
   \includegraphics[width=10cm]{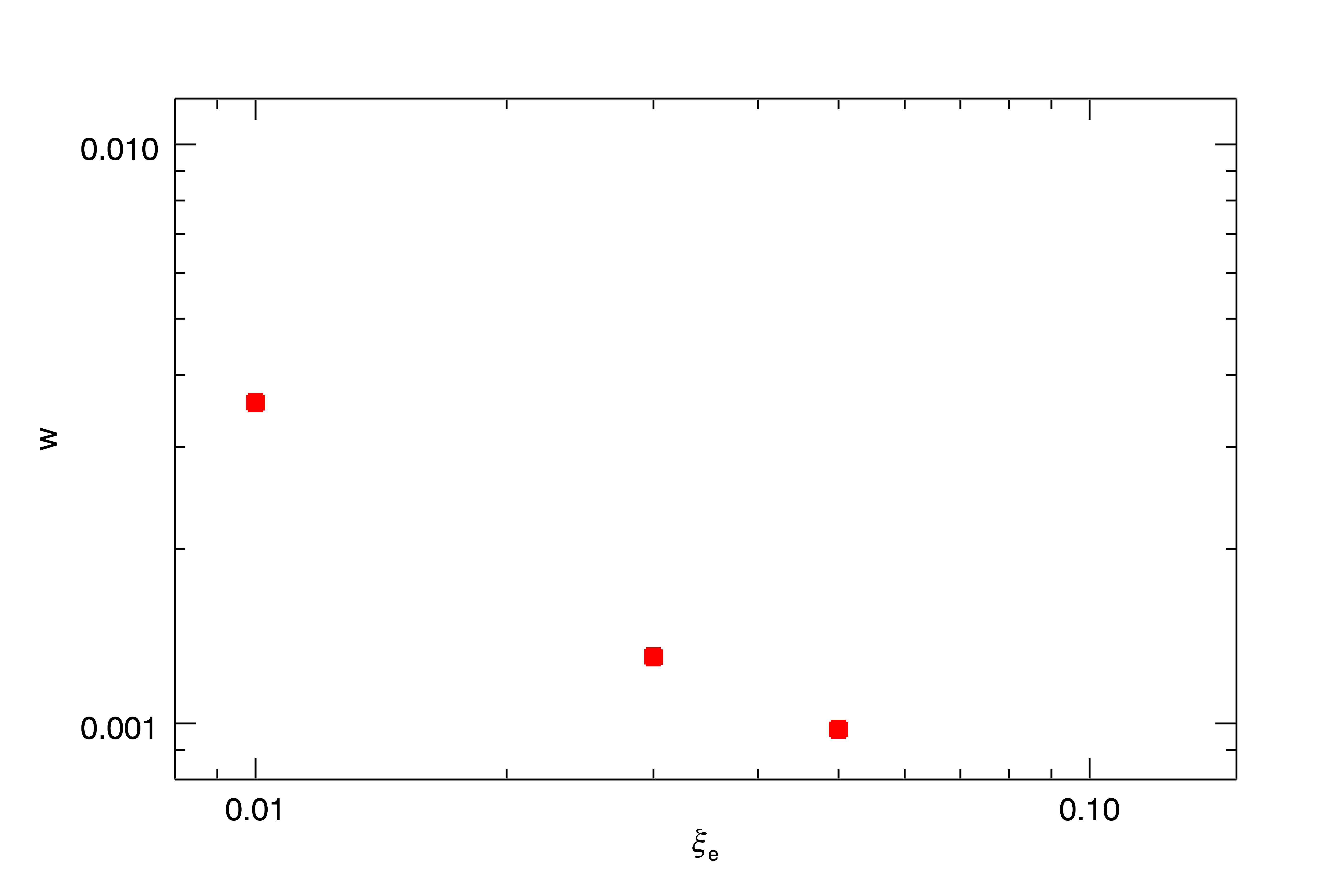}[h] 
   \caption{\small Transverse size of the emission region as a function of $\xi{\rm e}$.}
   \label{fig:width}
\end{figure}

In order to make more quantitative this result, in Fig. \ref{fig:width} we plot the radial width $w$ of the emission region at the point where the value is half the maximum as a function of $\xi_{\rm e}$. We observe a progressive decrease of the width for increasing values of the energy transferred to the non-thermal component. For the largest $\xi_{\rm e}=0.05$ the width is of the order of $10^{-3} z_0$, slightly smaller then the estimate of \citet{Bromberg09}. Therefore, even in not extreme conditions, the focusing due to the recollimation is so effective that the brightest region of the jet is as small as $\approx 7\times 10^{14}$ cm, not far from the size estimated for the rapid variability observed in 1222+216 \citep{Aleksic11, Tavecchio11}.

\begin{figure}[h!]
   \centering
   \includegraphics[width=6cm]{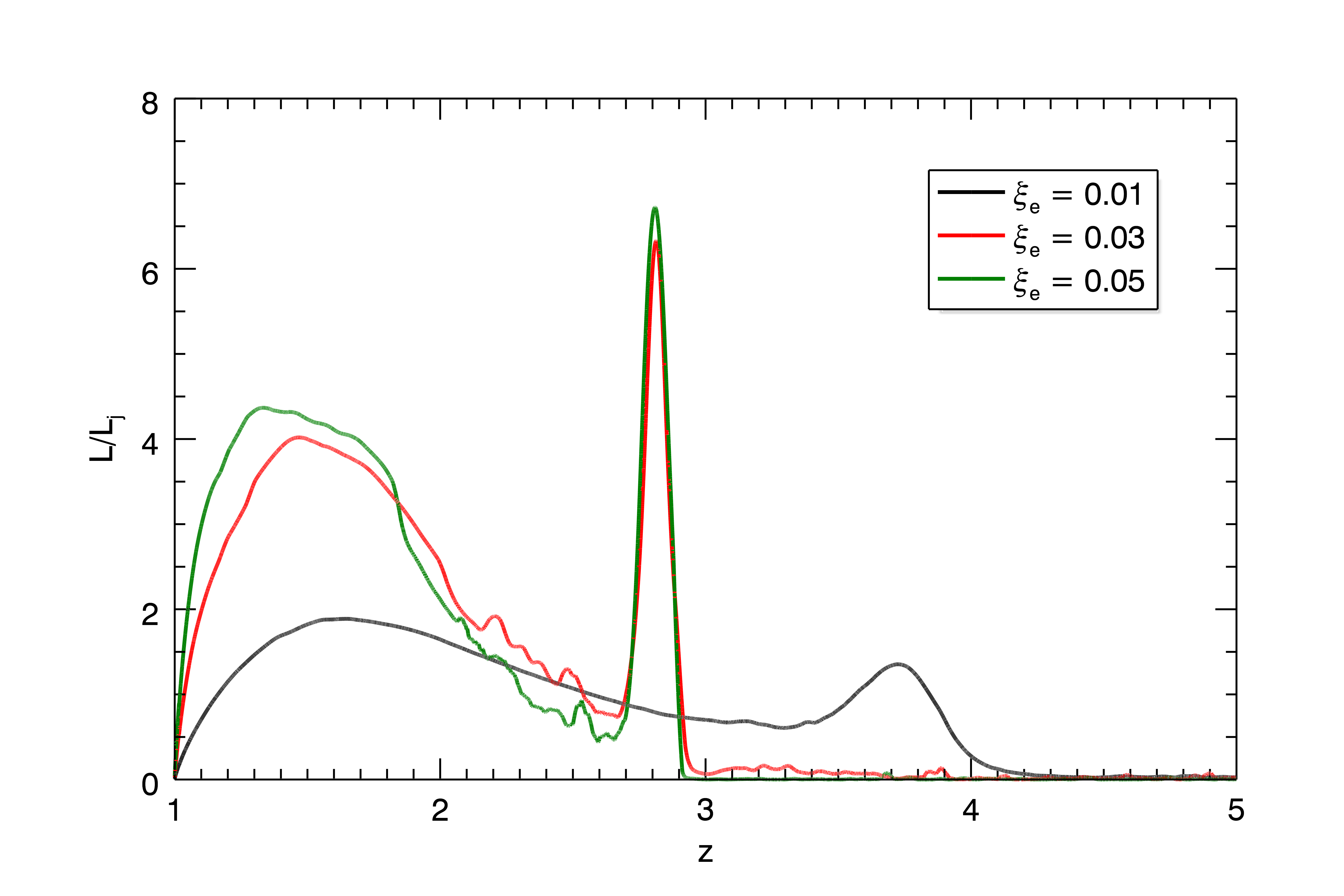} 
   \includegraphics[width=6cm]{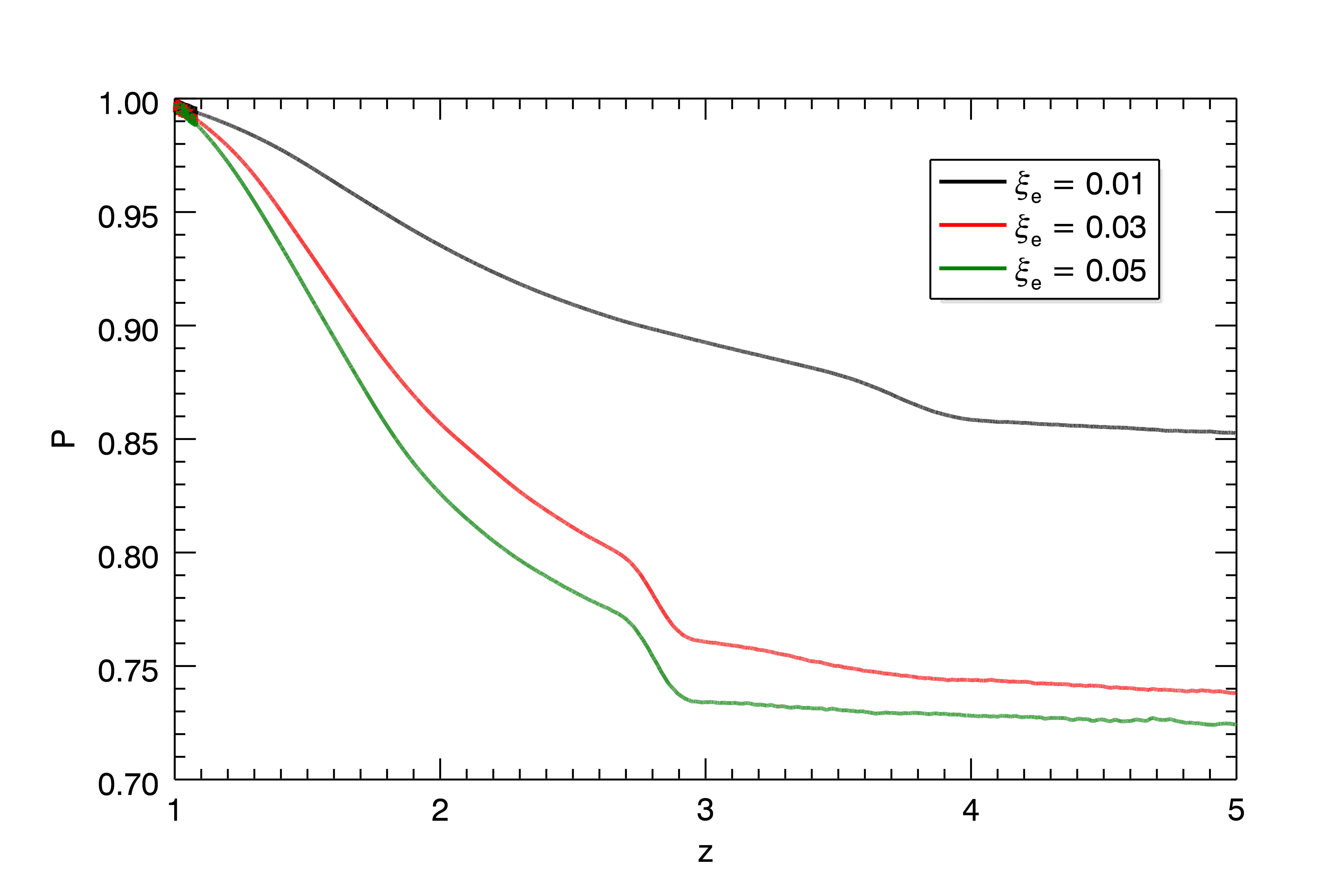} 
   \caption{\small  The left panel shows the observed emissivity integrated over planes normal to the $z$ axis as a function of the $z$ coordinate. The right panel shows the energy flux integrated  over planes normal to the $z$ axis as a function of the $z$ coordinate. }
   \label{fig:profiles}
\end{figure}

For illustration, we also report in Fig. \ref{fig:profiles} the profiles (integrated in planes normal to the jet axis) along the jet axis of the observed emissivity (left panel)
\begin{equation}\label{eq:emis}
L(z) = 2 \pi  \int_0^{r_D}  \epsilon _{\rm obs} (r,z) r dr ,
\end{equation}
where $\epsilon _{\rm obs}$ is defined in Eq. (\ref{eq:emisobs}) and of the jet energy flux (right panel)
\begin{equation}
P(z) = 2 \pi \int_0^{r_D} w (r, z)  \gamma^2 (r, z) v_z(r,z) r dr,
\end{equation}
for the three values of $\xi_{\rm e}$. Both quantities are normalized to the  jet power. The overall shape of the integrated emissivity $L(z)$ (left panel) is the same for all values of the losses, displaying a broad maximum around $z_0\sim 1.5$ in correspondence of the onset of the recollimation shock and a second, narrower maximum marking the region of minimum transverse size. For the lowest value of $\xi_{\rm e}=10^{-2}$ (black line) this second peak is relatively broad and not very high. For larger values of $\xi_{\rm e}$ (red and green lines) the overall emissivity increases and the second maximum becomes quite narrow, flagging the compactness of the emission region. Note that, in all cases, the area under the two peaks is of the same order, indicating that the total received luminosity comprises similar contribution from the large (hence slowly varying) recollimation shock and the compact (rapidly variable) region. The profiles of the energy flux (right panel) show the increasing importance of energy losses for increasing values of $\xi_{\rm e}$. After a smooth decrease, marking the progressive development of the recollimation shock, the curves (especially the two with $\xi_{\rm e}=0.03$ and 0.05) display a sudden jump, corresponding to the increased losses close to the region of smaller radius. After the expansion produced by the reflected shock the energy flux stays approximately constant. The fraction of the jet energy flux lost through the emitted radiation goes from $15\%$ for $\xi_{\rm e}=0.01$ to about a fraction $30\%$ for  $\xi_{\rm e}=0.05$. These values are  in agreement with the results of BL09. 

It is also interesting to consider the efficiency $\epsilon_{\rm diss}$ of the dissipation of the kinetic flux of the jet
that we define, similarly to  \citet{Nalewajko09}, as $L_{\rm kin}=\int \rho c^2 v_z \gamma_{\rm j} (\gamma_{\rm j} -1)$, where the integral is performed on planes normal to the $z$ direction. Comparing the values of $L_{kin}$ at injection and after the recollimation region, 
we derive $\epsilon_{\rm diss}=18\%$ for $\xi_{\rm e}=0.01$ and $\epsilon_{\rm diss}=35\%$ for $\xi_{\rm e}=0.05$. These values  are larger than the average shock dissipation efficiency derived by \citet{Nalewajko09}, that found values lower than 10\% for $\gamma_j\theta_j=1$. The reason for this higher dissipation efficiency should be looked for in the important radiation losses of the system, not included in the Nalewajko \& Sikora treatment. In fact, radiation losses lead to a lower value of the gas pressure downstream the recollimation shock. In order to maintain pressure balance with the external material, the system has thus to increase the fraction of jet luminosity dissipated into pressure. This is indeed naturally achieved through a narrower profile of the recollimation shock which increases the dissipation because of the larger component of the velocity normal to the shock.

\subsection{Evolution of perturbations and variability}

In the previous section we have studied in detail the stationary equilibrium structure of the recollimation shock. For astrophysical applications is important to have a glimpse on the temporal behavior of the structure and its properties when the jet is subject to perturbations. 

\begin{figure*}[t]
   \centering
   \includegraphics[width=4cm]{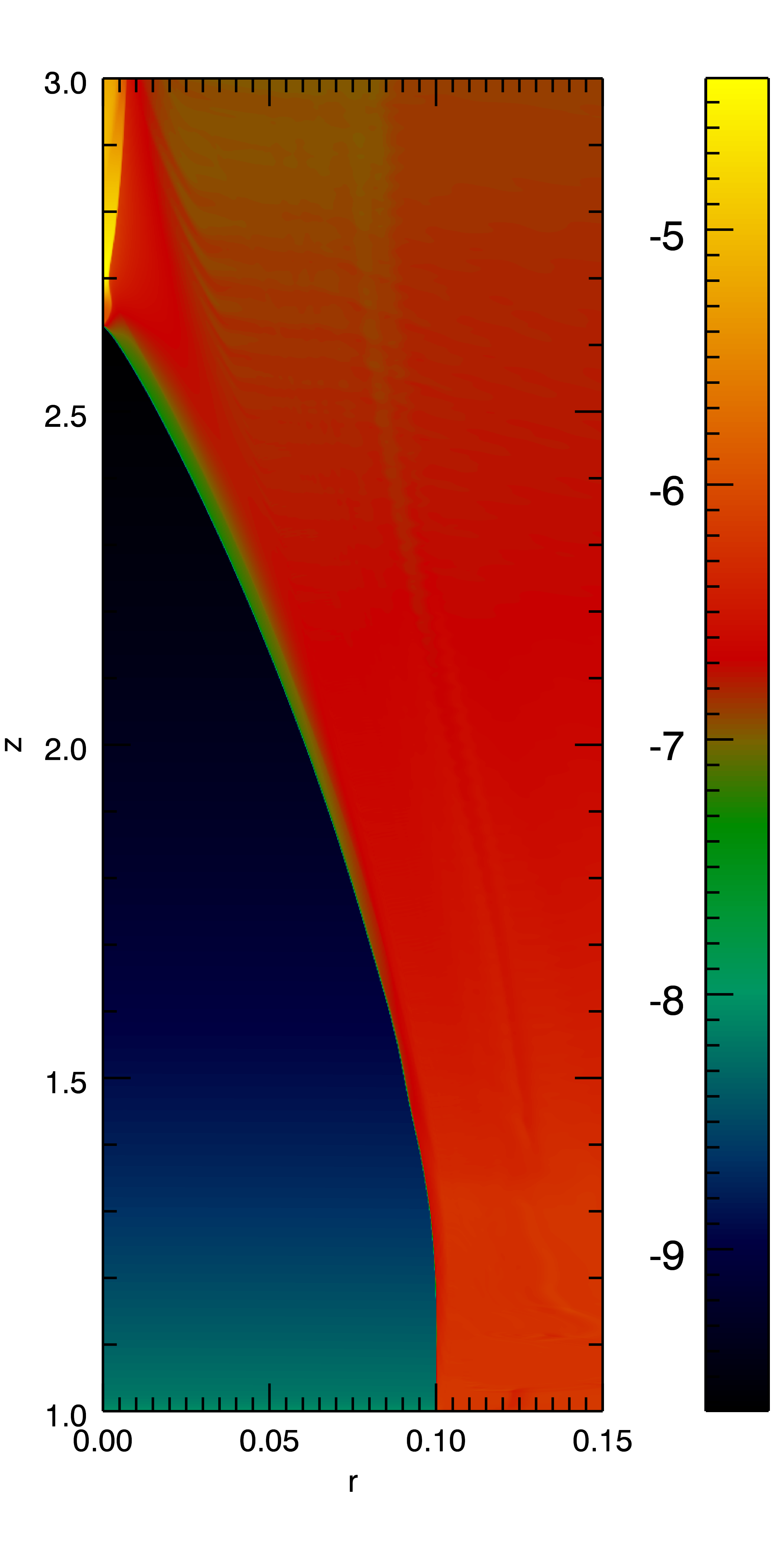} 
   \includegraphics[width=4cm]{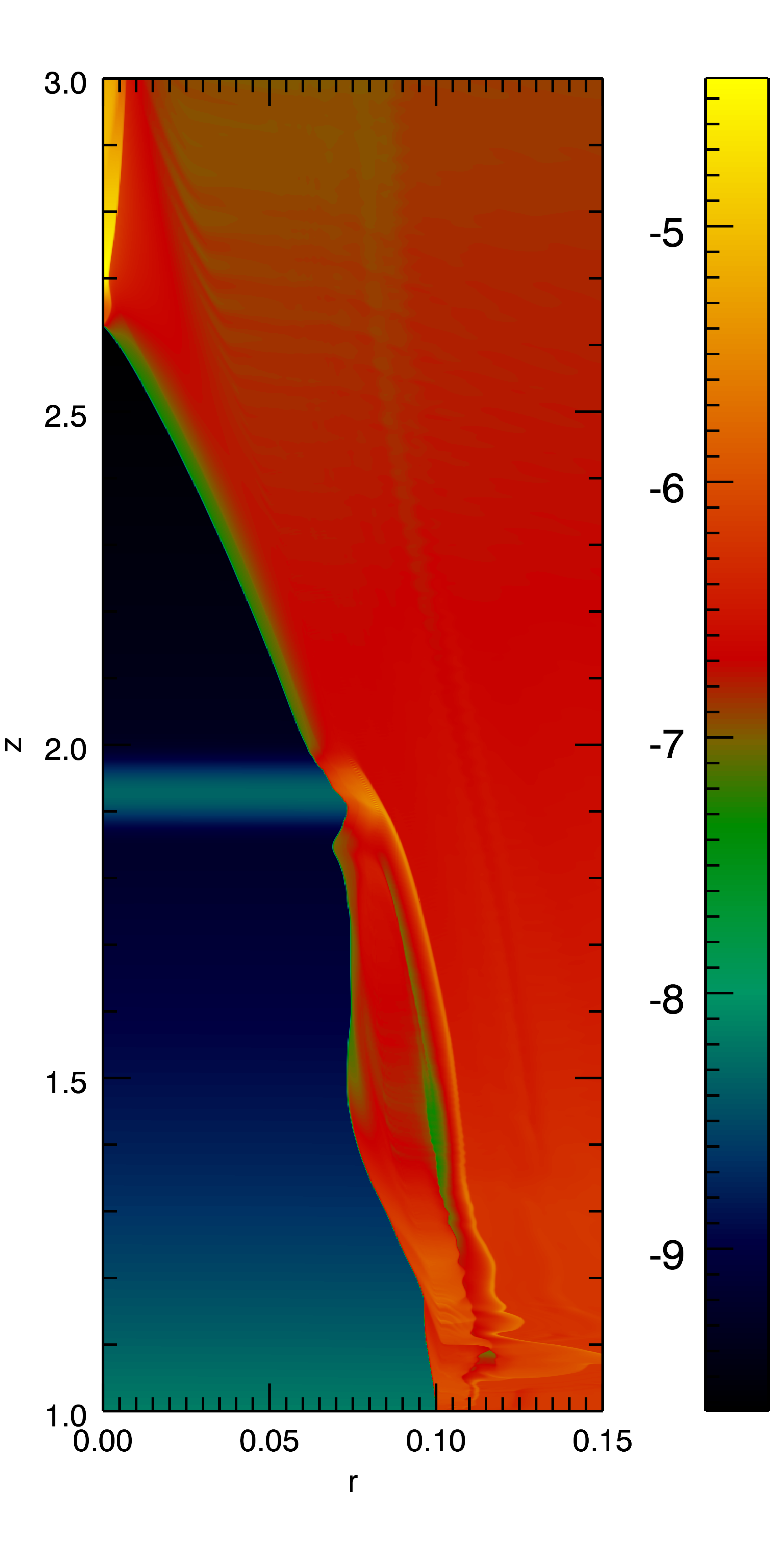} 
      \includegraphics[width=4cm]{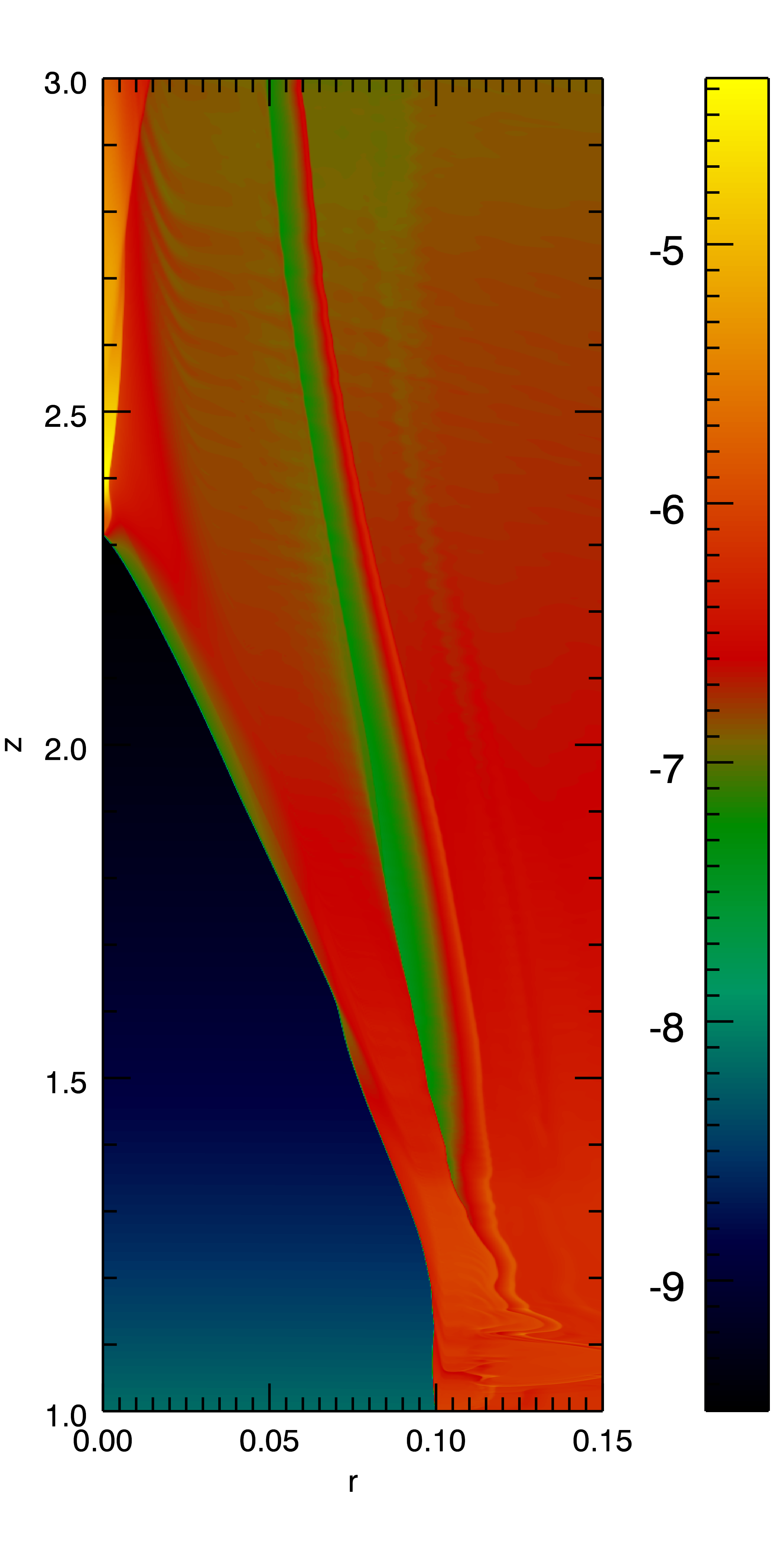} 
   \caption{\small  Pressure distribution (logarithmic scale) for the simulation including a propagating density perturbation. The three panel shows the pressure in the stationary (equilibrium) state, after a time $t = 0.9 z_0/c$ and, on the left, at $t=5z_0/c$. The figure clearly show the high pressure wake induced by the perturbation in the jet. In the last panel it is clearly visible the inward shift of the recollimation nozzle.}
\label{fig:pres_pert}
\end{figure*}

In order to study the variability behavior, we have injected a perturbation at the jet inlet and followed its evolution as it propagates along the jet. We choose to change the jet density, keeping a constant temperature, with a perturbation of the form:
\begin{equation}
\rho = \rho_0 \left\{1+ \epsilon \, \exp\left[-\left(\frac{t-t_0}{\tau}\right)^2\right]\right\}
\end{equation}
where $\epsilon$ is the relative amplitude of the perturbation, $\tau$ is the temporal width of the perturbation and $t_0$ is the time at which the perturbation reaches its  maximum value. From now on we take $t_0$ as the origin of time. For $\tau$ we took the value 0.05$z_0/c$ and we explored different value of $\epsilon$. Our results show that only quite large values of $\epsilon$ give appreciable effects on the observed emissivity, for example $\epsilon = 1$ gives only a $10\%$ increase in the emissivity, therefore we will discuss the results obtained for $\epsilon = 10$. In Figure \ref{fig:pres_pert} we show the logarithmic pressure distributions at three different times. In the left panel  we have the equilibrium distribution.  The middle panel is for $t = 0.9 z_0/c$ and we can see the propagating perturbation that leaves on the side an high pressure wake that compresses and modifies the structure of the recollimation shock. Finally the right panel is for $t=5z_0/c$, the perturbation is outside the region displayed in the figure, however its wake is still visible and  the recollimation shock is still deformed and the point at which the recollimation shock reaches the axis has moved inward.

\begin{figure*}[t]
   \centering
   \includegraphics[width=6cm]{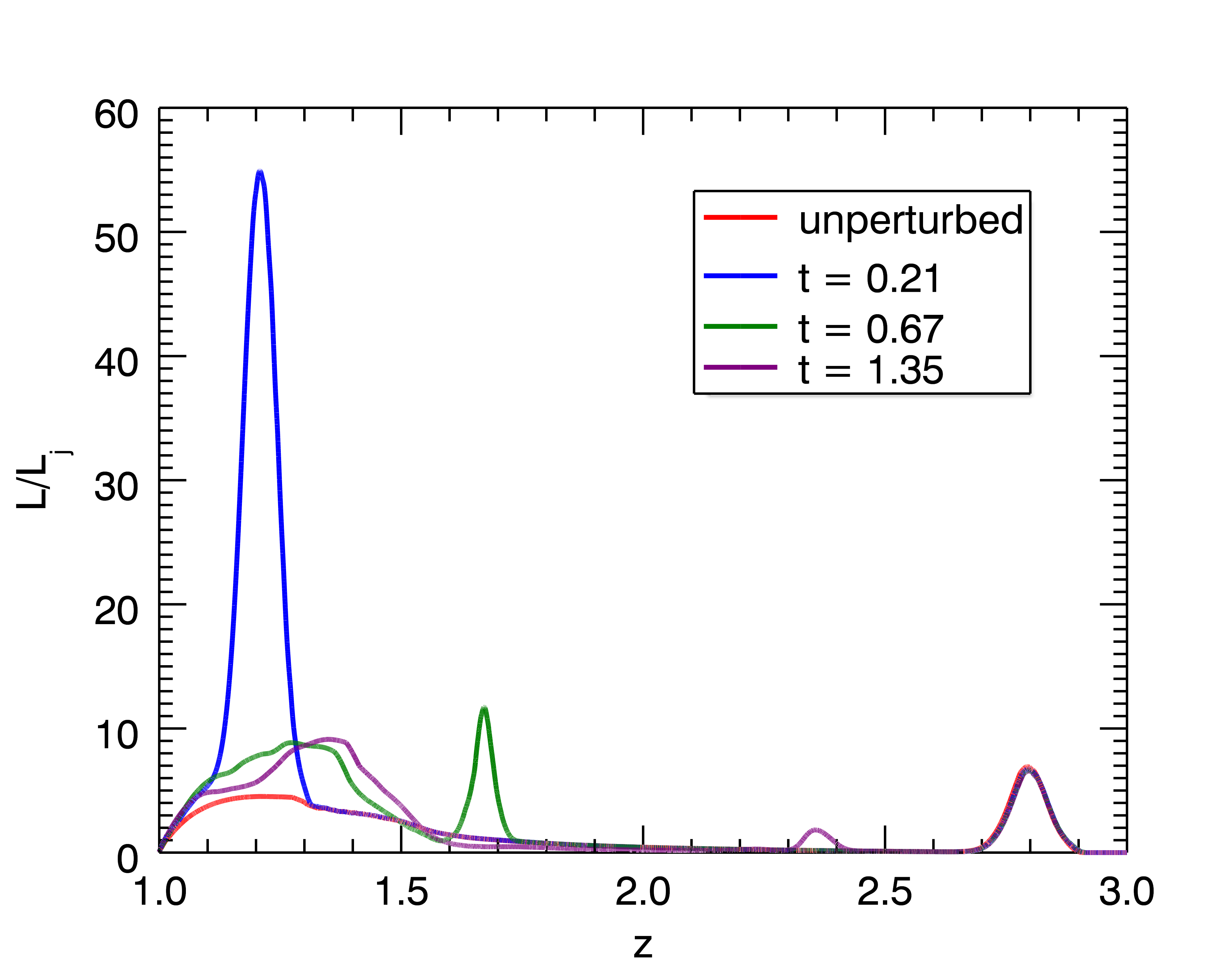} 
   \includegraphics[width=6cm]{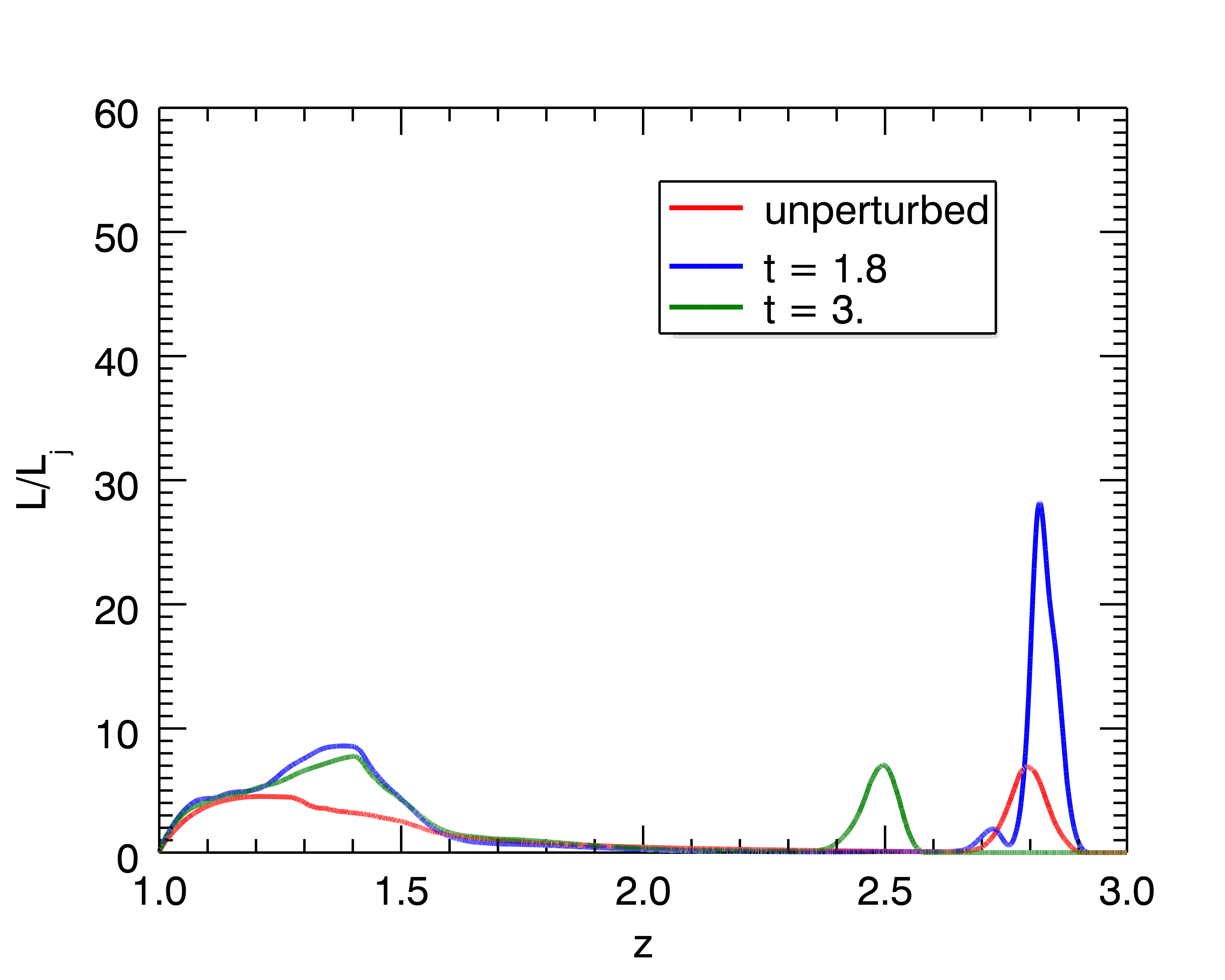} 
   \caption{\small  Profiles (integrated in planes normal to the jet axis) along the jet axis of the observed emissivity calculated for different times during the propagation of the perturbation. In both panels the red lines display the unperturbed profile. The left panel shows the profile at small times after the injection, while in the right panel we report the profiles at late times.}
\label{fig:emis_pert}
\end{figure*}

In Figure \ref{fig:emis_pert} we show  the profiles (integrated in planes normal to the jet axis) along the jet axis of the observed emissivity (Eq. \ref{eq:emisobs}) for different times. The left panel shows the first phases of propagation (corresponding to $t=0.21, 0.67 \; \hbox{and} \; 1.35 z_0/c$), with the red curve showing the unperturbed profile for comparison.  The peak corresponding to the perturbation decreases its amplitude as it moves outward. This decrease is related to the jet expansion. Because of the higher pressure and the shock deformation in the wake of the perturbation (see Fig.\ref{fig:pres_pert}), the inner jet maintains an emissivity that is approximately twice the unperturbed value even at larger times. In the right panel, the red curve shows always the unperturbed distribution for comparison, while the blue curve, corresponding to the time at which the perturbation reaches the recollimation region ($t=1.8z_0/c$), shows a burst of emissivity, with an observed luminosity increasing by a factor of $\approx 5$ with respect to the stationary case. Finally, the green curve in the right panel corresponds to a time when the perturbation has moved outside the displayed region, strongly decreasing its amplitude. In this case the peak in the emissivity at the recollimation point has moved inward, following the shift of the recollimation nozzle visible in the maps in Fig.  \ref{fig:pres_pert}.

These results, although obtained under rather idealized assumptions, outline a coherent scenario, qualitatively compatible with the behavior of powerful blazars during active states. In fact, the light curves of FSRQ during typical flare episodes show a long-term (weeks) increase of the flux (that in our simulations would be provided by the long duration active state of the perturbed inner jet), on top of which short timescale (days) variations are often detected. These rapid variations  could be related to the excitation of the recollimation nozzle by the incoming perturbation. In real astrophysical conditions we could also imagine a series of perturbations with different properties injected into the flow that would account for the erratic behavior of the real light curves.

%

\section{Discussion and Conclusions}
%
%
%
 
 We have developed a series of  axisymmetric simulations of a relativistic jet subject to recollimation caused by an external medium using the state-of-the-art code PLUTO and including the radiative losses mediated by a population of non-thermal electrons in the jet. The set up of the simulations is tailored to model jets associated to FSRQ, in particular the source PKS 1222+216, displaying rapid variability at VHE. Our simulations overcome several of the approximations introduced in the semi-analitical approach of \citet{Bromberg09}.
 
 The radiative cooling is then assumed to be dominated by the inverse Compton scattering of external radiation \citep{Sikora94, Ghisellini09}. The estimated effective timescale for  radiative losses, $\tau_c$, is much lower then the dynamical timescale $\tau_{\rm dyn}\sim z_o/c$, ensuring the effective cooling of the shocked plasma. In order to allow a sizeable energy loss by the jet we have also assumed that the emitting electrons, while cooling, are continuously heated, efficiently converting the energy dissipated by the shock in radiation. We have shown that, even with a relatively minor contribution of the non-thermal particles to the total pressure of the shocked material, $p_{\rm e}/p =\xi_{\rm e}= 0.05$, the fraction of energy flux lost by the flow can be as large as $30\%$, resulting in a significant deceleration of the flow downstream the recollimation zone.

The significant pressure loss caused by the radiative cooling allows the jet to be focused and extremely ``squeezed" by the external pressure. Consistently with the early findings by \citet{Bromberg09}, our simulations show that the minimum radius of the recollimation region can be as small as $10^{-3}z_0$, implying a light crossing time comparable to the timescale observed for PKS 1222+216. We have also produced maps for the observed emissivity of the flow, including the effects of the Doppler beaming. These shows that the emission is concentrated in a very thin elongated region close to the jet axis. We need to caution that the release of the axisymmetry constraint may lead to instabilities of this narrow region, whose outcome can however be  assessed only through three-dimensional simulations.

We have also explored the behavior of the recollimation shock under density perturbations injected at the base of the jet. Although highly idealized, these simulations show a rather complex phenomenology, that could be, at least qualitatively, related to the typical flaring pattern of powerful FSRQ.

We would like to remark that our calculations, similarly to those of BL09, should be considered somewhat optimistic. In fact the radiative losses induced in the system are maximized by the specific form of the cooling term adopted in the simulations, simply proportional to the thermal gas pressure. As already noted in Sect. 2, such prescription implicitly assumes that the emitting electrons are continuously heated by some unspecified mechanism, allowing to radiate an amount of energy larger than that put into the non-thermal component just after the shock front. As discussed by BL09, possible heating mechanisms include non-linear oscillations of the contact discontinuity surface or acceleration by shear flows. The investigation of this crucial aspects could be the subjects of future specific simulations.

\section*{Acknowledgments}
This work has been partly funded by a PRIN-INAF 2014 grant, the PRIN INAF SKA-CTA grant and a PRIN MIUR 2015. We acknowledge the CINECA award under the ISCRA initiative, for the availability of high performance computing resources.

\label{lastpage}

\begin{thebibliography}{34}
\expandafter\ifx\csname natexlab\endcsname\relax\def\natexlab#1{#1}\fi

\bibitem[{{Ackermann} {et~al.}(2016){Ackermann}, {Anantua}, {Asano}, {Baldini},
  {Barbiellini}, {Bastieri}, {Becerra Gonzalez}, {Bellazzini}, {Bissaldi},
  {Blandford}, {Bloom}, {Bonino}, {Bottacini}, {Bruel}, {Buehler}, {Caliandro},
  {Cameron}, {Caragiulo}, {Caraveo}, {Cavazzuti}, {Cecchi}, {Cheung}, {Chiang},
  {Chiaro}, {Ciprini}, {Cohen-Tanugi}, {Costanza}, {Cutini}, {D'Ammando}, {de
  Palma}, {Desiante}, {Digel}, {Di Lalla}, {Di Mauro}, {Di Venere}, {Drell},
  {Favuzzi}, {Fegan}, {Ferrara}, {Fukazawa}, {Funk}, {Fusco}, {Gargano},
  {Gasparrini}, {Giglietto}, {Giordano}, {Giroletti}, {Grenier}, {Guillemot},
  {Guiriec}, {Hayashida}, {Hays}, {Horan}, {J{\'o}hannesson}, {Kensei},
  {Kocevski}, {Kuss}, {La Mura}, {Larsson}, {Latronico}, {Li}, {Longo},
  {Loparco}, {Lott}, {Lovellette}, {Lubrano}, {Madejski}, {Magill}, {Maldera},
  {Manfreda}, {Mayer}, {Mazziotta}, {Michelson}, {Mirabal}, {Mizuno},
  {Monzani}, {Morselli}, {Moskalenko}, {Nalewajko}, {Negro}, {Nuss}, {Ohsugi},
  {Orlando}, {Paneque}, {Perkins}, {Pesce-Rollins}, {Piron}, {Pivato},
  {Porter}, {Principe}, {Rando}, {Razzano}, {Razzaque}, {Reimer}, {Scargle},
  {Sgr{\`o}}, {Sikora}, {Simone}, {Siskind}, {Spada}, {Spinelli}, {Stawarz},
  {Thayer}, {Thompson}, {Torres}, {Troja}, {Uchiyama}, {Yuan}, \&
  {Zimmer}}]{Ackermann16}
{Ackermann}, M., {Anantua}, R., {Asano}, K., {et~al.} 2016, \apjl, 824, L20

\bibitem[{{Aharonian} {et~al.}(2007){Aharonian}, {Akhperjanian}, {Bazer-Bachi},
  {Behera}, {Beilicke}, {Benbow}, {Berge}, {Bernl{\"o}hr}, {Boisson}, {Bolz},
  {Borrel}, {Boutelier}, {Braun}, {Brion}, {Brown}, {B{\"u}hler},
  {B{\"u}sching}, {Bulik}, {Carrigan}, {Chadwick}, {Clapson}, {Chounet},
  {Coignet}, {Cornils}, {Costamante}, {Degrange}, {Dickinson},
  {Djannati-Ata{\"i}}, {Domainko}, {Drury}, {Dubus}, {Dyks}, {Egberts},
  {Emmanoulopoulos}, {Espigat}, {Farnier}, {Feinstein}, {Fiasson},
  {F{\"o}rster}, {Fontaine}, {Funk}, {Funk}, {F{\"u}{\ss}ling}, {Gallant},
  {Giebels}, {Glicenstein}, {Gl{\"u}ck}, {Goret}, {Hadjichristidis}, {Hauser},
  {Hauser}, {Heinzelmann}, {Henri}, {Hermann}, {Hinton}, {Hoffmann}, {Hofmann},
  {Holleran}, {Hoppe}, {Horns}, {Jacholkowska}, {de Jager}, {Kendziorra},
  {Kerschhaggl}, {Kh{\'e}lifi}, {Komin}, {Kosack}, {Lamanna}, {Latham}, {Le
  Gallou}, {Lemi{\`e}re}, {Lemoine-Goumard}, {Lenain}, {Lohse}, {Martin},
  {Martineau-Huynh}, {Marcowith}, {Masterson}, {Maurin}, {McComb}, {Moderski},
  {Moulin}, {de Naurois}, {Nedbal}, {Nolan}, {Olive}, {Orford}, {Osborne},
  {Ostrowski}, {Panter}, {Pedaletti}, {Pelletier}, {Petrucci}, {Pita},
  {P{\"u}hlhofer}, {Punch}, {Ranchon}, {Raubenheimer}, {Raue}, {Rayner},
  {Renaud}, {Ripken}, {Rob}, {Rolland}, {Rosier-Lees}, {Rowell}, {Rudak},
  {Ruppel}, {Sahakian}, {Santangelo}, {Saug{\'e}}, {Schlenker}, {Schlickeiser},
  {Schr{\"o}der}, {Schwanke}, {Schwarzburg}, {Schwemmer}, {Shalchi}, {Sol},
  {Spangler}, {Stawarz}, {Steenkamp}, {Stegmann}, {Superina}, {Tam},
  {Tavernet}, {Terrier}, {van Eldik}, {Vasileiadis}, {Venter}, {Vialle},
  {Vincent}, {Vivier}, {V{\"o}lk}, {Volpe}, {Wagner}, {Ward}, \&
  {Zdziarski}}]{Aharonian07}
{Aharonian}, F., {Akhperjanian}, A.~G., {Bazer-Bachi}, A.~R., {et~al.} 2007,
  \apjl, 664, L71

\bibitem[{{Aharonian} {et~al.}(2017){Aharonian}, {Barkov}, \&
  {Khangulyan}}]{Aharonian17}
{Aharonian}, F.~A., {Barkov}, M.~V., \& {Khangulyan}, D. 2017, \apj, 841, 61

\bibitem[{{Albert} {et~al.}(2007){Albert}, {Aliu}, {Anderhub}, {Antoranz},
  {Armada}, {Baixeras}, {Barrio}, {Bartko}, {Bastieri}, {Becker}, {Bednarek},
  {Berger}, {Bigongiari}, {Biland}, {Bock}, {Bordas}, {Bosch-Ramon}, {Bretz},
  {Britvitch}, {Camara}, {Carmona}, {Chilingarian}, {Coarasa}, {Commichau},
  {Contreras}, {Cortina}, {Costado}, {Curtef}, {Danielyan}, {Dazzi}, {De
  Angelis}, {Delgado}, {de los Reyes}, {De Lotto}, {Domingo-Santamar{\'{\i}}a},
  {Dorner}, {Doro}, {Errando}, {Fagiolini}, {Ferenc}, {Fern{\'a}ndez}, {Firpo},
  {Flix}, {Fonseca}, {Font}, {Fuchs}, {Galante}, {Garc{\'{\i}}a-L{\'o}pez},
  {Garczarczyk}, {Gaug}, {Giller}, {Goebel}, {Hakobyan}, {Hayashida},
  {Hengstebeck}, {Herrero}, {H{\"o}hne}, {Hose}, {Hrupec}, {Hsu}, {Jacon},
  {Jogler}, {Kosyra}, {Kranich}, {Kritzer}, {Laille}, {Lindfors}, {Lombardi},
  {Longo}, {L{\'o}pez}, {L{\'o}pez}, {Lorenz}, {Majumdar}, {Maneva},
  {Mannheim}, {Mansutti}, {Mariotti}, {Mart{\'{\i}}nez}, {Mazin}, {Merck},
  {Meucci}, {Meyer}, {Miranda}, {Mirzoyan}, {Mizobuchi}, {Moralejo}, {Nieto},
  {Nilsson}, {Ninkovic}, {O{\~n}a-Wilhelmi}, {Otte}, {Oya}, {Paneque},
  {Panniello}, {Paoletti}, {Paredes}, {Pasanen}, {Pascoli}, {Pauss}, {Pegna},
  {Persic}, {Peruzzo}, {Piccioli}, {Prandini}, {Puchades}, {Raymers}, {Rhode},
  {Rib{\'o}}, {Rico}, {Rissi}, {Robert}, {R{\"u}gamer}, {Saggion}, {Saito},
  {S{\'a}nchez}, {Sartori}, {Scalzotto}, {Scapin}, {Schmitt}, {Schweizer},
  {Shayduk}, {Shinozaki}, {Shore}, {Sidro}, {Sillanp{\"a}{\"a}}, {Sobczynska},
  {Stamerra}, {Stark}, {Takalo}, {Tavecchio}, {Temnikov}, {Tescaro}, {Teshima},
  {Torres}, {Turini}, {Vankov}, {Vitale}, {Wagner}, {Wibig}, {Wittek},
  {Zandanel}, {Zanin}, \& {Zapatero}}]{Albert07}
{Albert}, J., {Aliu}, E., {Anderhub}, H., {et~al.} 2007, \apj, 669, 862

\bibitem[{{Aleksi{\'c}} {et~al.}(2014){Aleksi{\'c}}, {Antonelli}, {Antoranz},
  {Babic}, {Barres de Almeida}, {Barrio}, {Becerra Gonz{\'a}lez}, {Bednarek},
  {Berger}, {Bernardini}, {Biland}, {Blanch}, {Bock}, {Boller}, {Bonnefoy},
  {Bonnoli}, {Borla Tridon}, {Borracci}, {Bretz}, {Carmona}, {Carosi}, {Carreto
  Fidalgo}, {Colin}, {Colombo}, {Contreras}, {Cortina}, {Cossio}, {Covino}, {Da
  Vela}, {Dazzi}, {De Angelis}, {De Caneva}, {Delgado Mendez}, {De Lotto},
  {Doert}, {Dom{\'{\i}}nguez}, {Dominis Prester}, {Dorner}, {Doro},
  {Eisenacher}, {Elsaesser}, {Farina}, {Ferenc}, {Fonseca}, {Font}, {Fruck},
  {Garc{\'{\i}}a L{\'o}pez}, {Garczarczyk}, {Garrido Terrats}, {Gaug},
  {Giavitto}, {Godinovi{\'c}}, {Gonz{\'a}lez Mu{\~n}oz}, {Gozzini}, {Hadamek},
  {Hadasch}, {H{\"a}fner}, {Herrero}, {Hose}, {Hrupec}, {Idec}, {Kadenius},
  {Knoetig}, {Kr{\"a}henb{\"u}hl}, {Krause}, {Kushida}, {La Barbera}, {Lelas},
  {Lewandowska}, {Lindfors}, {Lombardi}, {L{\'o}pez-Coto}, {L{\'o}pez},
  {L{\'o}pez-Oramas}, {Lorenz}, {Lozano}, {Makariev}, {Mallot}, {Maneva},
  {Mankuzhiyil}, {Mannheim}, {Maraschi}, {Marcote}, {Mariotti},
  {Mart{\'{\i}}nez}, {Masbou}, {Mazin}, {Meucci}, {Miranda}, {Mirzoyan},
  {Mold{\'o}n}, {Moralejo}, {Munar-Adrover}, {Nakajima}, {Niedzwiecki},
  {Nilsson}, {Nowak}, {Orito}, {Overkemping}, {Paiano}, {Palatiello},
  {Paneque}, {Paoletti}, {Paredes}, {Partini}, {Persic}, {Prada}, {Prada
  Moroni}, {Prandini}, {Preziuso}, {Puljak}, {Reichardt}, {Reinthal}, {Rhode},
  {Rib{\'o}}, {Rico}, {R{\"u}gamer}, {Saggion}, {Saito}, {Saito}, {Salvati},
  {Satalecka}, {Scalzotto}, {Scapin}, {Schultz}, {Schweizer}, {Shore},
  {Sillanp{\"a}{\"a}}, {Sitarek}, {Snidaric}, {Sobczynska}, {Spanier}, {Spiro},
  {Stamatescu}, {Stamerra}, {Steinke}, {Storz}, {Sun}, {Suri{\'c}}, {Takalo},
  {Takami}, {Tavecchio}, {Temnikov}, {Terzi{\'c}}, {Tescaro}, {Teshima},
  {Thaele}, {Tibolla}, {Torres}, {Toyama}, {Treves}, {Uellenbeck}, {Vogler},
  {Wagner}, {Weitzel}, {Zandanel}, {Zanin}, \& {MAGIC
  Collaboration}}]{Aleksic14}
{Aleksi{\'c}}, J., {Antonelli}, L.~A., {Antoranz}, P., {et~al.} 2014, \aap,
  563, A91

\bibitem[{{Aleksi{\'c}} {et~al.}(2011){Aleksi{\'c}}, {Antonelli}, {Antoranz},
  {Backes}, {Barrio}, {Bastieri}, {Becerra Gonz{\'a}lez}, {Bednarek},
  {Berdyugin}, {Berger}, {Bernardini}, {Biland}, {Blanch}, {Bock}, {Boller},
  {Bonnoli}, {Borla Tridon}, {Braun}, {Bretz}, {Ca{\~n}ellas}, {Carmona},
  {Carosi}, {Colin}, {Colombo}, {Contreras}, {Cortina}, {Cossio}, {Covino},
  {Dazzi}, {de Angelis}, {de Cea Del Pozo}, {de Lotto}, {Delgado Mendez},
  {Diago Ortega}, {Doert}, {Dom{\'{\i}}nguez}, {Dominis Prester}, {Dorner},
  {Doro}, {Elsaesser}, {Ferenc}, {Fonseca}, {Font}, {Fruck}, {Garc{\'{\i}}a
  L{\'o}pez}, {Garczarczyk}, {Garrido}, {Giavitto}, {Godinovi{\'c}}, {Hadasch},
  {H{\"a}fner}, {Herrero}, {Hildebrand}, {Hose}, {Hrupec}, {Huber}, {Jogler},
  {Klepser}, {Kr{\"a}henb{\"u}hl}, {Krause}, {La Barbera}, {Lelas}, {Leonardo},
  {Lindfors}, {Lombardi}, {L{\'o}pez}, {Lorenz}, {Majumdar}, {Makariev},
  {Maneva}, {Mankuzhiyil}, {Mannheim}, {Maraschi}, {Mariotti},
  {Mart{\'{\i}}nez}, {Mazin}, {Meucci}, {Miranda}, {Mirzoyan}, {Miyamoto},
  {Mold{\'o}n}, {Moralejo}, {Nieto}, {Nilsson}, {Orito}, {Oya}, {Paoletti},
  {Pardo}, {Paredes}, {Partini}, {Pasanen}, {Pauss}, {Perez-Torres}, {Persic},
  {Peruzzo}, {Pilia}, {Pochon}, {Prada}, {Prada Moroni}, {Prandini}, {Puljak},
  {Reichardt}, {Reinthal}, {Rhode}, {Rib{\'o}}, {Rico}, {R{\"u}gamer},
  {R{\"u}ger}, {Saggion}, {Saito}, {Saito}, {Salvati}, {Satalecka},
  {Scalzotto}, {Scapin}, {Schultz}, {Schweizer}, {Shayduk}, {Shore},
  {Sillanp{\"a}{\"a}}, {Sitarek}, {Sobczynska}, {Spanier}, {Spiro}, {Stamerra},
  {Steinke}, {Storz}, {Strah}, {Suri{\'c}}, {Takalo}, {Tavecchio}, {Temnikov},
  {Terzi{\'c}}, {Tescaro}, {Teshima}, {Thom}, {Tibolla}, {Torres}, {Treves},
  {Vankov}, {Vogler}, {Wagner}, {Weitzel}, {Zabalza}, {Zandanel}, \&
  {Zanin}}]{Aleksic11}
{Aleksi{\'c}}, J., {Antonelli}, L.~A., {Antoranz}, P., {et~al.} 2011, \aap,
  530, A4

\bibitem[{{Aller} {et~al.}(1985){Aller}, {Aller}, \& {Hughes}}]{Aller85}
{Aller}, H.~D., {Aller}, M.~F., \& {Hughes}, P.~A. 1985, \apj, 298, 296

\bibitem[{{Arlen} {et~al.}(2013){Arlen}, {Aune}, {Beilicke}, {Benbow},
  {Bouvier}, {Buckley}, {Bugaev}, {Cesarini}, {Ciupik}, {Connolly}, {Cui},
  {Dickherber}, {Dumm}, {Errando}, {Falcone}, {Federici}, {Feng}, {Finley},
  {Finnegan}, {Fortson}, {Furniss}, {Galante}, {Gall}, {Griffin}, {Grube},
  {Gyuk}, {Hanna}, {Holder}, {Humensky}, {Kaaret}, {Karlsson}, {Kertzman},
  {Khassen}, {Kieda}, {Krawczynski}, {Krennrich}, {Maier}, {Moriarty},
  {Mukherjee}, {Nelson}, {O'Faol{\'a}in de Bhr{\'o}ithe}, {Ong}, {Orr}, {Park},
  {Perkins}, {Pichel}, {Pohl}, {Prokoph}, {Quinn}, {Ragan}, {Reyes},
  {Reynolds}, {Roache}, {Saxon}, {Schroedter}, {Sembroski}, {Staszak},
  {Telezhinsky}, {Te{\v s}i{\'c}}, {Theiling}, {Tsurusaki}, {Varlotta},
  {Vincent}, {Wakely}, {Weekes}, {Weinstein}, {Welsing}, {Williams}, {Zitzer},
  {VERITAS Collaboration}, {Jorstad}, {MacDonald}, {Marscher}, {Smith},
  {Walker}, {Hovatta}, {Richards}, {Max-Moerbeck}, {Readhead}, {Lister},
  {Kovalev}, {Pushkarev}, {Gurwell}, {L{\"a}hteenm{\"a}ki}, {Nieppola},
  {Tornikoski}, \& {J{\"a}rvel{\"a}}}]{Arlen13}
{Arlen}, T., {Aune}, T., {Beilicke}, M., {et~al.} 2013, \apj, 762, 92

\bibitem[{{Begelman} {et~al.}(2008){Begelman}, {Fabian}, \&
  {Rees}}]{Begelman08}
{Begelman}, M.~C., {Fabian}, A.~C., \& {Rees}, M.~J. 2008, \mnras, 384, L19

\bibitem[{{Bromberg} \& {Levinson}(2007)}]{Bromberg07}
{Bromberg}, O. \& {Levinson}, A. 2007, \apj, 671, 678

\bibitem[{{Bromberg} \& {Levinson}(2009)}]{Bromberg09}
{Bromberg}, O. \& {Levinson}, A. 2009, \apj, 699, 1274

\bibitem[{{Dermer}(1995)}]{Dermer95}
{Dermer}, C.~D. 1995, \apjl, 446, L63

\bibitem[{{Ghisellini} {et~al.}(1998){Ghisellini}, {Celotti}, {Fossati},
  {Maraschi}, \& {Comastri}}]{Ghisellini98}
{Ghisellini}, G., {Celotti}, A., {Fossati}, G., {Maraschi}, L., \& {Comastri},
  A. 1998, \mnras, 301, 451

\bibitem[{{Ghisellini} \& {Tavecchio}(2008)}]{Ghisellini08}
{Ghisellini}, G. \& {Tavecchio}, F. 2008, \mnras, 386, L28

\bibitem[{{Ghisellini} \& {Tavecchio}(2009)}]{Ghisellini09}
{Ghisellini}, G. \& {Tavecchio}, F. 2009, \mnras, 397, 985

\bibitem[{{Ghisellini} \& {Tavecchio}(2010)}]{Ghisellini10}
{Ghisellini}, G. \& {Tavecchio}, F. 2010, \mnras, 409, L79

\bibitem[{{Giannios}(2013)}]{Giannios13}
{Giannios}, D. 2013, \mnras, 431, 355

\bibitem[{{Komissarov} \& {Falle}(1998)}]{Komissarov98}
{Komissarov}, S.~S. \& {Falle}, S.~A.~E.~G. 1998, \mnras, 297, 1087

\bibitem[{{Liu} \& {Bai}(2006)}]{Liu06}
{Liu}, H.~T. \& {Bai}, J.~M. 2006, \apj, 653, 1089

\bibitem[{{Marscher}(2014)}]{Marscher14}
{Marscher}, A.~P. 2014, \apj, 780, 87

\bibitem[{{Mignone} \& {Bodo}(2005)}]{Mignone05a}
{Mignone}, A. \& {Bodo}, G. 2005, \mnras, 364, 126

\bibitem[{{Mignone} {et~al.}(2007){Mignone}, {Bodo}, {Massaglia}, {Matsakos},
  {Tesileanu}, {Zanni}, \& {Ferrari}}]{PLUTO}
{Mignone}, A., {Bodo}, G., {Massaglia}, S., {et~al.} 2007, \apjs, 170, 228

\bibitem[{{Mignone} {et~al.}(2005){Mignone}, {Plewa}, \& {Bodo}}]{Mignone05b}
{Mignone}, A., {Plewa}, T., \& {Bodo}, G. 2005, \apjs, 160, 199

\bibitem[{{Mignone} {et~al.}(2012){Mignone}, {Zanni}, {Tzeferacos}, {van
  Straalen}, {Colella}, \& {Bodo}}]{AMRPLUTO}
{Mignone}, A., {Zanni}, C., {Tzeferacos}, P., {et~al.} 2012, \apjs, 198, 7

\bibitem[{{Nalewajko} \& {Sikora}(2009)}]{Nalewajko09}
{Nalewajko}, K. \& {Sikora}, M. 2009, \mnras, 392, 1205

\bibitem[{{Narayan} \& {Piran}(2012)}]{Narayan12}
{Narayan}, R. \& {Piran}, T. 2012, \mnras, 420, 604

\bibitem[{{Petropoulou} {et~al.}(2016){Petropoulou}, {Giannios}, \&
  {Sironi}}]{Petropoulou16}
{Petropoulou}, M., {Giannios}, D., \& {Sironi}, L. 2016, \mnras, 462, 3325

\bibitem[{{Romero} {et~al.}(2017){Romero}, {Boettcher}, {Markoff}, \&
  {Tavecchio}}]{Romero17}
{Romero}, G.~E., {Boettcher}, M., {Markoff}, S., \& {Tavecchio}, F. 2017, \ssr,
  207, 5

\bibitem[{{Sikora} {et~al.}(1994){Sikora}, {Begelman}, \& {Rees}}]{Sikora94}
{Sikora}, M., {Begelman}, M.~C., \& {Rees}, M.~J. 1994, \apj, 421, 153

\bibitem[{{Sokolov} {et~al.}(2004){Sokolov}, {Marscher}, \&
  {McHardy}}]{Sokolov04}
{Sokolov}, A., {Marscher}, A.~P., \& {McHardy}, I.~M. 2004, \apj, 613, 725

\bibitem[{{Stawarz} {et~al.}(2006){Stawarz}, {Aharonian}, {Kataoka},
  {Ostrowski}, {Siemiginowska}, \& {Sikora}}]{Stawarz06}
{Stawarz}, {\L}., {Aharonian}, F., {Kataoka}, J., {et~al.} 2006, \mnras, 370,
  981

\bibitem[{{Tavecchio} {et~al.}(2011){Tavecchio}, {Becerra-Gonzalez},
  {Ghisellini}, {Stamerra}, {Bonnoli}, {Foschini}, \& {Maraschi}}]{Tavecchio11}
{Tavecchio}, F., {Becerra-Gonzalez}, J., {Ghisellini}, G., {et~al.} 2011, \aap,
  534, A86

\bibitem[{{Urry} \& {Padovani}(1995)}]{Urry95}
{Urry}, C.~M. \& {Padovani}, P. 1995, \pasp, 107, 803

\bibitem[{{Vovk} \& {Babi{\'c}}(2015)}]{Vovk15}
{Vovk}, I. \& {Babi{\'c}}, A. 2015, \aap, 578, A92

\end{thebibliography}
\end{document}